\documentclass[namedreferences]{SolarPhysics}
\usepackage[optionalrh]{spr-sola-addons} 
\usepackage{epsfig}
\usepackage{graphicx}
\usepackage{color}
\usepackage{url}

\begin{document}

\begin{article}

\begin{opening}

\title{Search for Short-Term Periodicities in the Sun's Surface 
Rotation: A Revisit}

\author{J.\ Javaraiah$^1$\sep R.K. Ulrich$^2$\sep L. Bertello$^2$\sep 
J.E.~Boyden$^2$}

 \institute{(1)\ Indian Institute of Astrophysics, Bangalore-560 034, India.\\
email: \url{jj@iiap.res.in}\\
}

 \institute{(2)\ Department of Physics and Astronomy, UCLA,
 430 Portola Plaza,   Los Angeles, CA 90095-1547,
 USA\\ 
email: \url{ulrich@astro.ucla.edu}\\
}

\runningauthor{J. Javaraiah \it et al.}
\runningtitle{Periodicities in the Sun's Surface Rotation}

\begin{abstract}
We have used the daily values of the equatorial rotation rate determined from
the Mt. Wilson daily  Doppler velocity measurements  during
 the period  3 December 1985\,--\,5 March 2007 to search 
for periodicities  in the solar equatorial rotation rate
  on time-scales shorter than 11 years. 
After the daily values  have been
binned into  61-day intervals, 
a cosine fit with a period of one year was applied to the
sequence to remove a seasonal trend. The spectral
properties of this sequence were
then investigated using a standard Fourier analysis, the
maximum-entropy methods, and the Morlet-wavelet analysis. 
From the analysis of the Fourier power spectrum we detected
peaks with periodicities around 7.6 years, 2.8 years,  
1.47 years, 245 days, 182 days, and 158 days, but none of them
were  at a statistically significant level.
In the  Morlet-wavelet analysis  the $\approx$ 1.47 year periodicity 
is  detected only for 1990, $i.e.$ near the  maximum 
of cycle~22, and near the end of cycle~22 in 1995.
From the same wavelet analysis
we found some evidence for the existence of a 2.8-year periodicity and a 245-day  
   periodicity in the equatorial rotation  rate  around the years
1990 and 1992,  respectively.
In the data taken during the period  1996\,--\,2007, when
the Mt. Wilson spectrograph instrumentation was more stable, we were
not able to detect any signal from the wavelet analysis.
Thus, the detected  
periodicities  during the period before the year 1996
 could be  artifacts of 
 frequent changes in the Mt. Wilson spectrograph instrumentation.
 However,
the temporal behavior of most of the activity phenomena  
during cycles 22 (1986\,--\,1996) and 23 
(after 1997) is considerably different. Therefore,  
the presence of the aforementioned short-term periodicities
 during the last cycle and absence of them in the current cycle may, 
in principle, be real 
 temporal behavior of the solar   
 rotation during these cycles.
\end{abstract}

\end{opening}

\section{Introduction}
Temporal variations in the Sun occur on 
many time scales. The time scales relevant for dynamical process range from minutes (lifetime of small
convective elements such as granules) to years/decades/centuries (solar cycles)
 to billion of years (evolution). 
Studies of short-term variations in solar activity and their solar 
cycle dependence may  greatly help 
 for better understanding the basic process of solar activity cycle  
and for predicting the level of solar 
activity~\cite{ks02,ksb05,ao06,jj07,jj08,fb07,os07}.   
 The current model  
is that the solar dynamo, thought to be responsible for the solar 
magnetic-activity cycle, operates near 
the base of the convection zone near the interface between convection and 
radiation zones~\cite{rw92,oss03}. This so-called Tachocline  is a layer of
 strong radial shear 
where the solar rotation profile changes from having a latitude
dependency in the convection zone to a pure radial profile in the 
radiative zone~\cite{sz92}. The variation of solar rotation with the 11-year 
 solar cycle is well-established by now.   
The 11-year torsional oscillations were
 discovered by~\inlinecite{hl80},
using Mt. Wilson Doppler velocity measurements, and have been 
 confirmed using different data sets and methods. 
Helioseismic observations show that the torsional 
oscillations are not just
a superficial phenomenon but they extend to at least the upper 
third of the solar convection zone~\cite{howe00a}. 
 Variations  on the few other time scales
in the coefficients of solar differential 
rotation~\cite{jg95,jg97,jj98,jj99,jj03,jj05,jk99,jbu05} 
and the residual rotation~\cite{brw06}, 
including 
one which is approximately equal to the Gleissberg cycle, 
 have been found using   sunspot group 
data.
However,
variations in the solar rotation on shorter time scales are more difficult
to determine because several observational 
and instrumental effects can produce spurious peaks with
similar periodicities.

\inlinecite{howe00b}
detected  a  1.3 year periodicity in the Sun's internal rotation 
rate near the base of the convection zone, 
using GONG and SOHO/MDI helioseismic data over the 
period May 1995 to November 1999. 
A similar  periodicity is  found in 
sunspot activity~\cite{ks02}  
 and in solar-wind speed~\cite{rich94},
  suggesting that this periodicity may be associated 
with the basic  process of the solar dynamo. In addition, 
a similar periodicity  
is also found in  the
interplanetary magnetic field~\cite{lock01} and 
many other related  solar and geomagnetic 
data, 
 suggesting that, possibly,  there exists a  
 direct connection between the basic mechanism of solar 
activity and that of interplanetary planetary magnetic
 field (for a recent and comprehensive review  see~\opencite{os07}). 
 Such a connection  may be possible  through the    
mediation of the solar wind~\cite{georg05}.
\inlinecite{jk99} analysed 
  the sun's surface ``mean'' rotation rate  
determined from the Mt. Wilson
  velocity data (1986\,--\,1994) and found,  beside a 
few other short-term periodicities,  
 a statistically significant 1.2 $\pm$ 0.2-year 
periodicity.  
This periodicity in the surface rotation data may be related to  
the 1.3-year periodicity  in the solar activity and the internal rotation.  
 However, \inlinecite{howe00b} did not find a 1.3-year periodicity
 in the sun's surface rotation rate determined from  
 the MDI and the GONG data. The aim of this paper is to improve 
 the analysis described in~\inlinecite{jk99} by
using relatively longer and better calibrated daily data of 
 the equatorial rotation  rate determined from the
Mt. Wilson  Doppler velocity measurements.
In addition, we investigated other short-term  
periodicities and their solar cycle dependence. 

In the next two sections we describe in some detail
the data and the analysis performed. 
The power spectra
of the time series were calculated using both the standard Fourier
analysis and the maximum entropy method (MEM). 
In addition, we have applied the Morlet wavelet analysis to study the 
solar-cycle dependence of these periodicities. In Section~4 we present 
 conclusions   
 and    briefly discuss  them.

\section{Data and Analysis}

It is standard practice to summarize solar differential rotation
profiles by fitting them to minimize least squares of the residuals 
 with the functional form:
$$\omega(\phi) = A + B \sin^2 \phi + C \sin^4 \phi ,\eqno(1) $$ 

\noindent where $\omega(\phi)$ is the solar angular velocity at 
latitude $\phi$, the coefficient  $A$  represents
the equatorial rotation rate and the coefficients $B$ and $C$  measure the 
latitudinal gradient in the rotation rate, 
with $B$ representing mainly low latitudes and $C$ representing largely
higher latitudes. This formula is chosen largely for historical reasons
but can be related to a spherical harmonic expansion.

\begin{figure}[ht]
\centerline{\includegraphics[width=11.0cm]{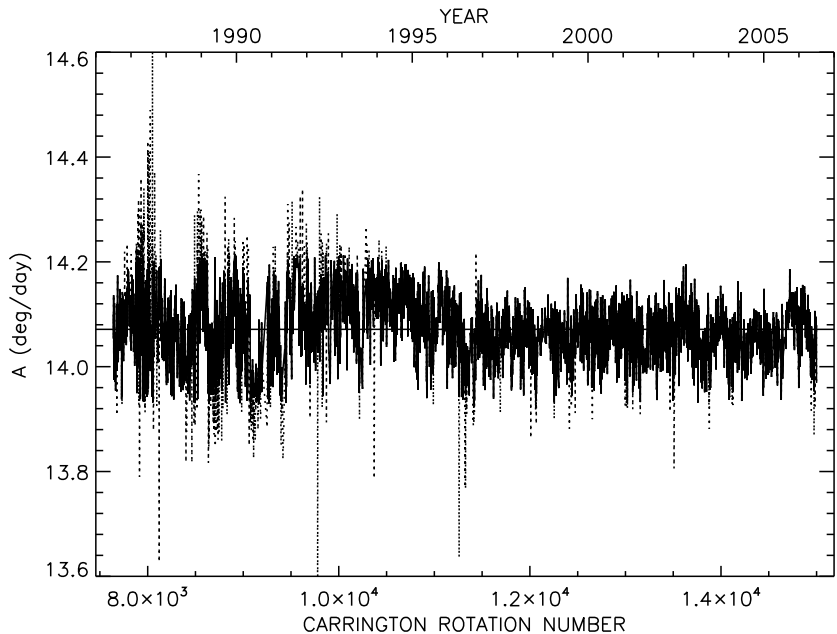}}
\caption{Variations  in $A$ derived from 
the Mt. Wilson daily  Doppler measurements during 3 December 1985 
to 5 March 2007.  
The  dotted curve represents the 
  uncorrected 
data, and the  solid curve represents the corrected data after 
 removal  
of the large spikes which are greater than 2$\sigma$, where 
$\sigma = 0.071$ deg/day is the standard deviation of the 
uncorrected data 
($\sigma = 0.056$ deg/day for the corrected data).
The 1$\sigma$ 
uncertainty of an individual
measurement is about 0.01 deg/day, much smaller than the 
aforementioned  $\sigma$ values  of the original 
and the corrected data.} 
\end{figure}

We have used the daily values of 
the equatorial rotation rate ($A$)
 derived from the Mount Wilson Doppler measurements 
 during the period 3 December 1985 to 5 March 2007.
This period covers  solar cycle~22 and most of cycle~23.
 In the present analysis we have used the data corrected for the 
 scattered-light effect (for a detail see~\opencite{ulrich01}) and
removed the 
   very large spikes ($viz.$ $> 2\sigma$, where $\sigma$ is the 
standard deviation of the original time series). 
Figure~1 shows  both the original (dotted curve)
 and the corrected (solid curve)  time series 
of $A$. The time series has data gaps which vary in size,
with a maximum gap of 49 days during Carrington rotation numbers
 1560\,--\,1608.  
In order to produce an uninterrupted time series, 
we have binned the above corrected daily data into 120 consecutive 61-day
intervals. The 61-day  interval was chosen in order to have a 
sufficient amount of data in the intervals near  
a  large gap. Figure~2 shows 
the 61-day binned time series of the sidereal $A$ values.
In the  same figure the horizontal  dotted lines indicate 
the rms deviation in the  rotation rate.
 The one-year periodicity is found in 
several  solar-activity indices, but its origin is doubtful. That is, 
it is difficult to rule out the possibility that this 
periodicity  is not due to  influence of seasonal effects. Therefore,   
we removed the 1.0 year periodicity  from the times series to determine
the short-term periodicities  in $A$ shown in Figure~2.
 Figure~3 shows the
corresponding time series after subtracting a cosine fit of 
1.0 year from the time series shown in Figure~2. 
From  the analysis of Figures~2 and 3 we can
 see that there are systematic 
variations in $A$ on  time 
scales of the order of a few years to few days. The amplitudes
of these variations 
are  significantly  
larger than  the  rms deviation  at some epochs, before the
year 1996.  

\begin{figure}[ht]
\centerline{\includegraphics[width=12.0cm]{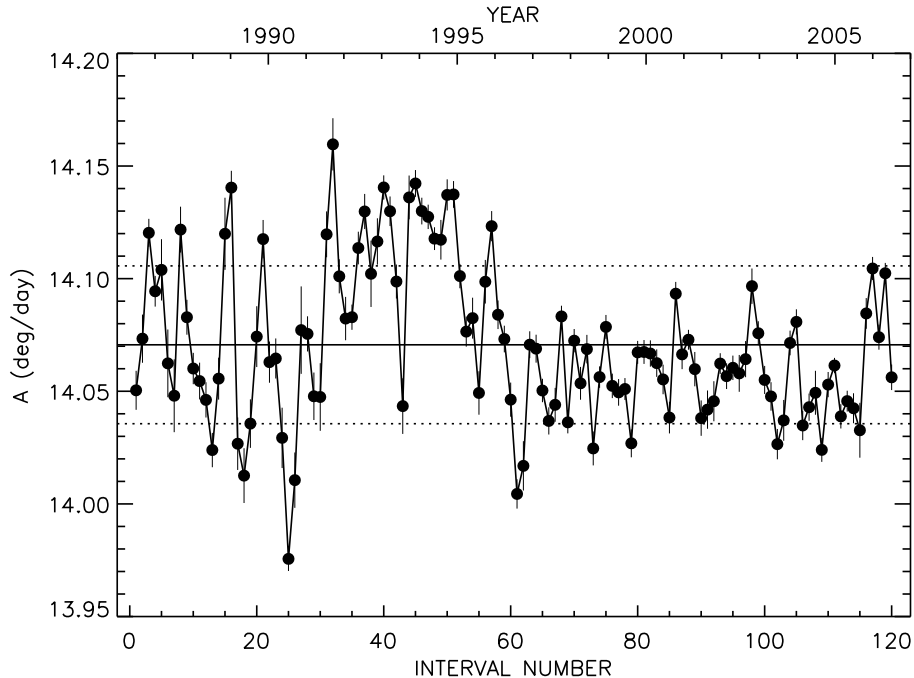}}
\caption{Variation of  $A$ 
in the 61-day  intervals of the corrected data obtained after removing 
the large spikes in the daily data. 
 The  horizontal solid line represents 
the mean and 
the  horizontal dotted lines indicate the corresponding rms deviation, 
 which is  about 0.035 deg/day.}   
\end{figure}

 As can be seen in Figure~1 there are some 
 large jumps  in the  
  original time series (dotted-curve)  of $A$. 
A few of  these large jumps 
are associated 
with the  time-intervals during which the following  
changes were made  in the Mount Wilson 
Spectrograph: 
 During 14 February 1989 to 2 May 1989, fiber-optics were rearranged for
 ``rubber-band'' grams.
 During 1 May 1990 to 4 May 1990, exit slit box was rebuilt with inch-worm and 
 exit slits were realigned.
On 4 September 1990,  a new grating box was installed.
On 16 November 1990,  entrance slit was made narrower and  spectrograph optics
 were realigned.
On 11 December 1991,  Littrow lenses were rotated to reduce astigmatism and reduced 
the entrance slit width again.
On 29 December 1996,  Littrow lenses were removed and cleaned.
In order  to implement this log-book information 
in  the analysis,   
we first  arbitrarily  
 removed some  high spikes  
around these time-intervals. 
But we found that it
increased the inconsistency  of the time  series,  $i.e.$  
it increased the number of  
 gaps in the data series and also the sizes of  some gaps 
are found to be increased considerably. Hence,  we  did not implement the  
deletions. 
Moreover,  many of those spikes are absent in the corrected time series
(the solid-curve of Figure~1)  
and the remaining ones are washed out in the aforementioned 61-day averages 
(Figures~2 and 3).

We have computed  the fast Fourier transform(FFT),  maximum-entropy 
method (MEM),  
 and Morlet wavelet  power spectra 
 of the data shown in Figures~2 and 3. 
The MEM FORTRAN program was provided to us by  A.V. Raveendran, 
 and was also used in the earlier papers~\cite{jg95,jg97}. 
We have used 
the wavelet program by~\inlinecite{torr98}.
The results of these analyses are described in the next section.

\begin{figure}[ht]
\centerline{\includegraphics[width=12.0cm]{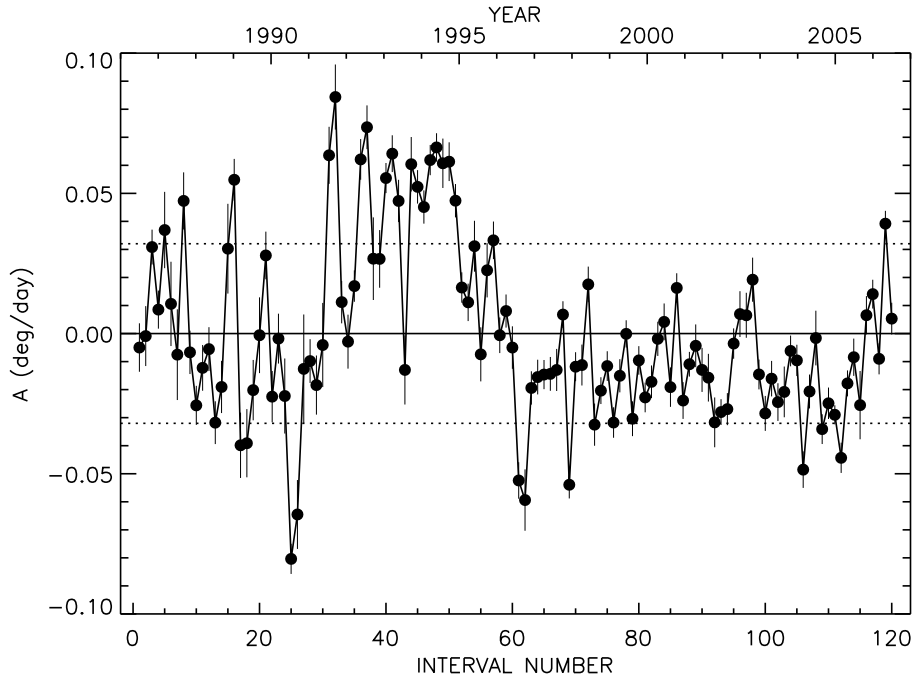}}
\caption{Variation of  $A$ 
in the 61-day  intervals of the corrected data obtained after removing 
the large spikes in the daily data and also  
 after subtracting a cosine fit  of 
1.0 year period.
 The  horizontal solid line represents 
the mean and  
 the  horizontal dotted lines indicate the corresponding rms deviation, 
 which is  about 0.032 deg/day.}   
\end{figure}

\section{Power Spectrum Analysis}
\subsection{\scriptsize FFT Analysis}
Figure~4  shows the FFT power spectrum    
of the variations in  $A$ shown in Figure~3.
Before computing the FFT,  the mean value 
of the series  was subtracted from the data and a cosine
bell function was applied to the first and the last 10\% of the time
series. Because of the requirement of the specific code we used for
the calculation of the Morlet wavelet and to keep the consistency between the
two analyses, we have padded the time series with eight zeros so that the
 number of data points (128) corresponds to an exact power of two.

The significance level of the peaks in a FFT power spectrum can be computed
by assuming that the mean power spectrum  can be  modeled using either a white-noise or 
a red-noise spectrum. For the case of a red-noise spectrum, the discrete Fourier 
power spectrum is given, after normalization, by 

$$p_k = \frac{1 - \alpha^2}{1 + \alpha^2 - 2\alpha \cos (2\pi k /N)}, \eqno$$
 
\noindent where $ k = 0, ..., N/2$ is the frequency index, $N$ is the number
 of data points, 
and $\alpha$ is the lag-one auto-correlation coefficient (when $\alpha = 0.0$ 
 we obtain the white-noise spectrum).  
 In this case, for a peak to be significant
at a given significant level, higher values of power are required compared 
to the white-noise background spectrum~\cite{torr98}.
We find    $\alpha = 0.62$ for the time series of 
$A$ shown in Figure~3. 
In the FFT power spectrum shown in Figure~4   there are 
 noticeable   peaks around the
 frequencies 21\,--\,7 year$^{-1}$, 
2.7 year$^{-1}$, 1.4 year$^{-1}$, 0.668 year$^{-1}$ (250 day period),  
and 0.411 year$^{-1}$ (150 day period).
For a white-noise model only the peaks at frequency 21\,--\,7  year$^{-1}$
 are above 3$\sigma$ level ($>$ 99\% confidence level).
From the red-noise model 
 we find  no peak is 
significant at a 90\% confidence 
level.  

\begin{figure}[ht]
\centerline{\includegraphics[width=12.0cm]{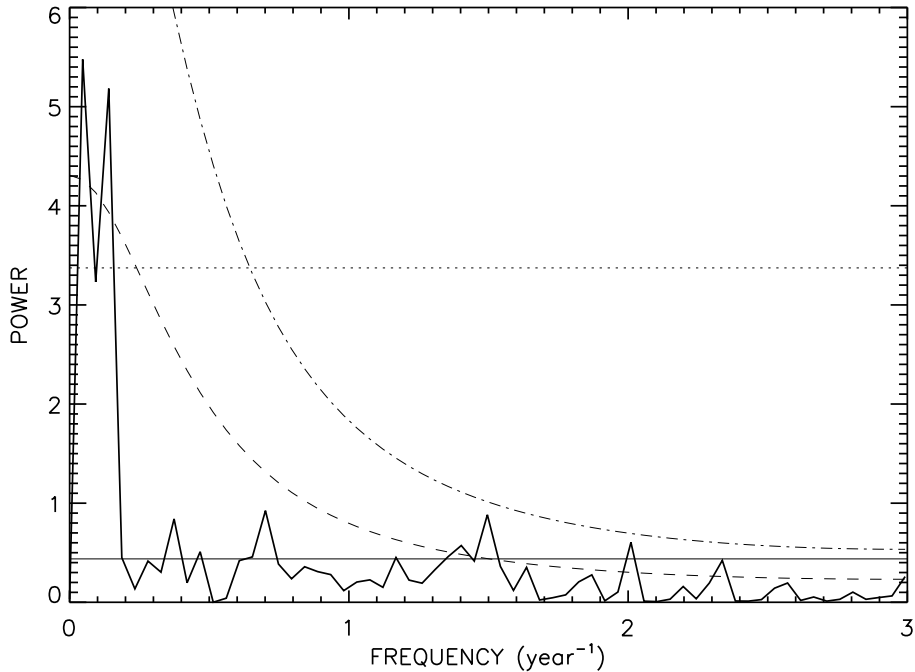}}
\caption{FFT power of spectrum of $A$  normalized by $N/(2s^2)$ 
where $s^2$ is the variance of the corresponding 
 time series shown in Figure~3. 
The  solid and the  dotted  horizontal lines represent the mean  and 
3$\sigma$ (from the mean) levels, respectively.  
The  dashed and the  dashed-dotted curves represent the 
corresponding 
 mean  and 90\% confidence  red-noise spectra. The corresponding   
 red-noise, $\alpha$, is  0.62.} 
\end{figure}

 We have also determined 
 the FFT power spectra of $A$ during the  
moving-time-intervals (MTI) successively shifted by 61 days and 365 days.   
We derived these MTI series from
 the data time series 
 shown in Figure~1,  in which the spikes     
above the $2\sigma$ level are removed. We did not show these spectra 
because we found no notable difference 
between these spectra and the ones shown in Figure~4, in the 
low-frequency side correspond to the aforementioned periodicities.

\subsection{\scriptsize MEM Analysis}

A different approach for determining the value of the periodicities 
in a short time series with higher accuracy is to compute
the power spectrum using a maximum entropy method (MEM).
An important step in this method is the optimum selection 
of the order $M$ of the autoregressive process.
If $M$ is chosen too low the spectrum is over-smoothed and the high-resolution 
potential is lost. If $M$ is chosen too high, frequency shifting and 
spontaneous splitting 
 of the spectral peaks occurs.  
In order to find the correct values of the periodicities-particularly 
the  $\approx$ 21\,--\,7 year periodicity seen in the FFT power spectrum, 
 we computed  
MEM power spectra choosing  various 
values for   $M$ in the range ($N/3$, $N/2$) 
as suggested by~\inlinecite{ulbis75}, 
where $N$ is the total number of intervals in the 
analyzed time series. 
We find that $M = N/3$ is  suitable in the present MEM 
analysis ($i.e.$, the peaks are sharp and well separated).
Figure~5 shows the MEM power  spectra of $A$. This spectrum  
shows  the values  
$\approx$ 7.6 year, 2.8 year,  
1.47 year, 245 day, 182 day,  and 158 day for the 
periodicities noticed  from the
FFT analysis.

\begin{figure}[ht]
\centerline{\includegraphics[width=12.0cm]{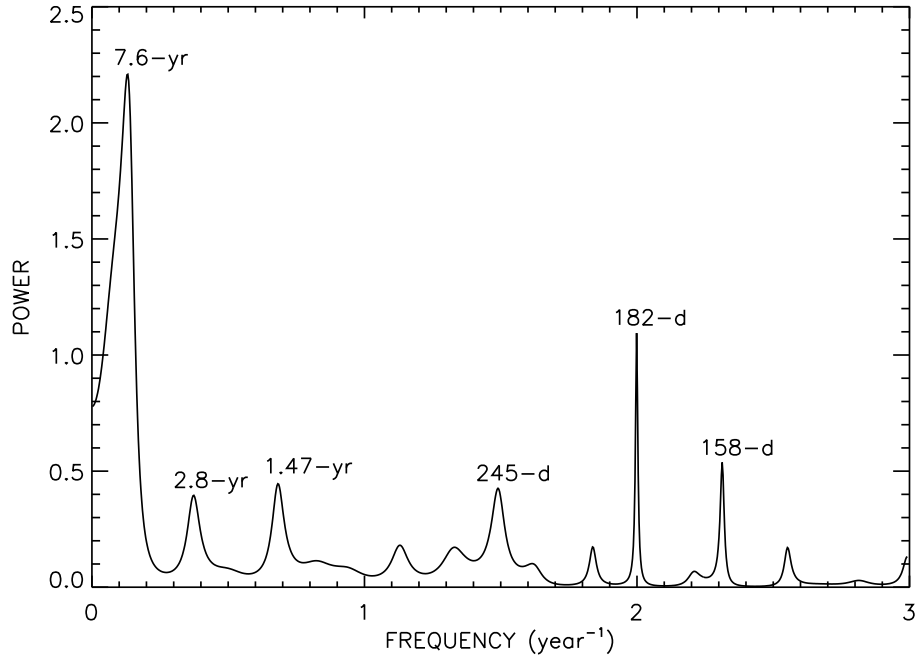}}
\caption{MEM power  spectrum  of $A$ correspond to the  
 corrected with one-year period filtered
 time series 
 (shown in Figure~3).  
Near  each considerably 
large peak, the value of the corresponding periodicity is mentioned.} 
\end{figure}

\subsection{\scriptsize Wavelet Analysis}
Wavelet analysis is a powerful method for 
analyzing localized variations 
of the power within a time series at many different 
frequencies~\cite{torr98}. 
The
 short-term periodicities 1.3 year, 150\,--\,160 day, etc. 
 in several solar activity 
phenomena have been analysed using this technique ($e.g.$, \opencite{obb98}; \opencite{ks02}; \opencite{prab02};
 \opencite{bal04};
 \opencite{ksb05}; \opencite{mvv06}; \opencite{cr06}; \opencite{fb07}). 
Hence, we used  the Morlet-wavelet analysis,  which also helps  
     to  determine  
the   solar-cycle dependence  of the short-term 
periodicities.   
Figure~6 shows the  wavelet spectra, 
normalized by $1/s^2$,
 of the corrected with one-year period filtered 
 time series of $A$ shown in Figure~3, 
 where $s^2$ is the variance of the same time series.
 The wavelet spectrum suggests that the 1.47-year and 
2.8-year 
 periodicities might have  
 occurred around the year 1990 (and 1995), and 
 the 245-day periodicity might have  
 occurred around 1992,
in the variation of A.  
These  periodicities are   above the 95\%    
confidence  level (corresponding to the  green contour level).
 The $\approx$ 7.6 year periodicity
is inside the  cone-of-influence (COI).     
Therefore, this periodicity is not detected here
unambiguously, although it 
 is above the 99\% confidence level in the FFT power spectrum (Figure~4). 
The available  data are inadequate   
 to accurately determine the correct value as well as 
the time dependence of this periodicity. On the other hand, 
such a periodicity seems present  in the rotation rates of magnetic features.
\inlinecite{sp85} have found  a seven-year periodicity in the 
rotation rate derived from Ca$^+$ K  plage data during the 
period 1951\,--\,1981.
The existence of an approximate eight-year periodicity is found 
in both the equatorial 
rotation rate and the latitudinal gradient   
of the rotation  determined from over 
100 years of Greenwich and Solar Optical Observatory Network (SOON) 
 sunspot group
 data~\cite{jk99,jj05}.

\begin{figure}[ht]
\centerline{\includegraphics[width=11.0cm]{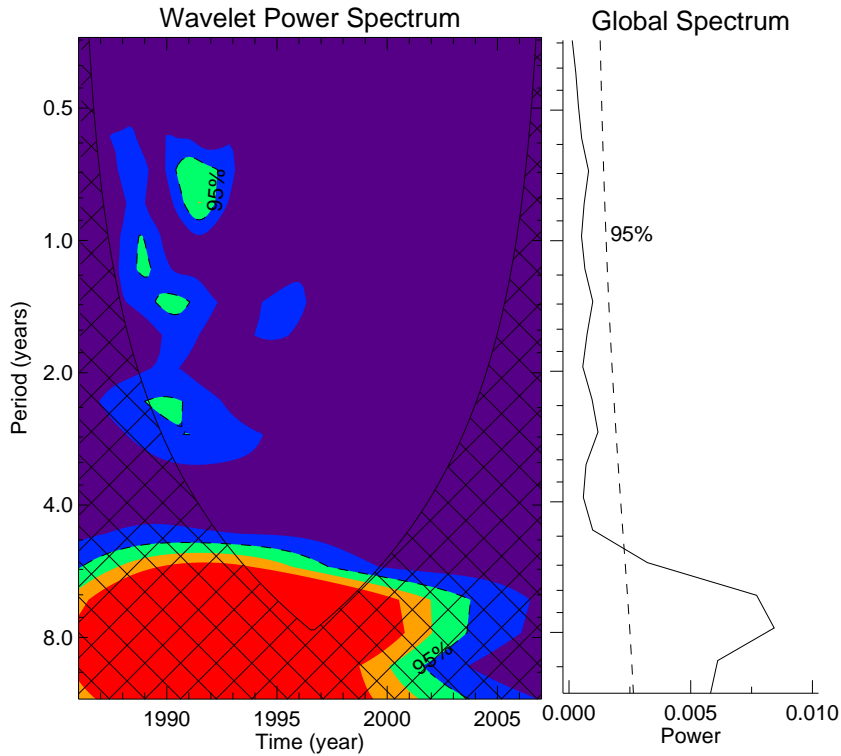}}
\caption{Wavelet power spectrum and the corresponding 
global power spectrum of $A$ using the Morlet wavelet, normalized 
by $1/s^2$, where $s^2$ is the variance of the corresponding time series
shown in Figure~3.
The contours  are at normalized variances of 1.5 (blue), 3.0 (green),
 4.5 (yellow) and 6.0 (red). The  dashed curve represent the
 95\% confidence level.
 Cross-hatched regions indicate the cone of 
influence, where edge effects become important (Torrence and Compo, 1998).}
\end{figure}

\section{Conclusion and Discussion}
The power spectral analyses of  the Sun's surface 
equatorial rotation rate 
  determined from  the 
Mt. Wilson daily  Doppler velocity measurements  during 
 the period  3 December 1985\,--\,5 March 2007  
suggests the existence of  
 7.6 year, 2.8 year,  
1.47 year, 245 day, 182 day and 158 day  
periodicities in the surface equatorial rotation rate
during the period  before  1996. 
 However,  there is no variation of any kind in the 
more accurately measured data during the period after  1995. 
That is, the aforementioned 
periodicities in the  data during the period before the year 1996
 may be  artifacts of the 
uncertainties 
of those data  due to the frequent changes in the instrumentation of 
the Mt. Wilson spectrograph.  Therefore, the results of the 
present analysis are largely consistent with result of  no 
variations  in the solar-surface equatorial rotation rate occur 
found by~\inlinecite{ulber96}.

On the other hand,   
the temporal
behavior of  most of the activity phenomena
is considerably different  during cycles 22  and 23. 
In several solar-activity phenomena the 1.3 year periodicity 
 has been found to be   
 dominant during cycle~22 and  weak or absent 
in the later period (see~\opencite{os07}). In addition, 
recently, \inlinecite{howe07} reported 
that the 1.3-year periodicity in the rotation rate at the 
base of the convection 
zone does  not persist after 2001.  
 Hence, presence of the short-term periodicities
 during  cycle~22 (1986\,--\,1996) and absence of them in cycle~23 (after 1996)
 may indicate that the 
 temporal behavior of the
 rotation is also considerably  different between these cycles.
Therefore, in spite of the fact that quality of the 
 data during the period 1986\,--\,1995 is poor,
 the  periodicities
 in $A$ found in this data  may  represent 
the corresponding real variations in the surface equatorial rotation
 rate during the period 1986\,--\,1995 and  need to be confirmed 
 from an independent data set.

\begin{acknowledgements}
We thank the anonymous referee for useful comments and suggestions. 
J.J. also thanks Dr. R.W. Komm for discussion and suggestions. 
J.J. did a major part of this
work at the department of Physics and Astronomy, UCLA. J.J.   
 gratefully acknowledges the funding 
by NSF grant ATM-0236682. 
Wavelet software was provided by 
 C. Torrence and G. Compo and  is 
available at \url{http//paos.colorado.edu/research/wavelets}. 
The MEM software was provided to J.J. by Dr. A.V. Raveendran. 
\end{acknowledgements}

\end{article}
\end{document}